\documentclass[fleqn,usenatbib]{mnras}

\usepackage{mathptmx}

\usepackage[T1]{fontenc}
\usepackage{ae,aecompl}

\usepackage{graphicx}	
\usepackage{amsmath}	
\usepackage{amssymb}	
\usepackage{bm}
\usepackage{color}
\usepackage[all]{hypcap}
\usepackage{hyperref}
\usepackage{multicol}
\usepackage{multirow}
\usepackage{flushend}
\usepackage{txfonts}
\usepackage{epstopdf}
\usepackage{float}
\usepackage{subcaption}

\usepackage{etoolbox}
\makeatletter
\patchcmd\@combinedblfloats{\box\@outputbox}{\unvbox\@outputbox}{}{\errmessage{\noexpand patch failed}}
\makeatother

\title[Jet inclination for TDE Swift J1644+57]{Estimation of the jet inclination angle for the TDE Swift J1644+57}

\author[S. Chakraborty et al.]{
Sudip Chakraborty$^{1}$\thanks{E-mail: sudip.chakraborty@tifr.res.in},
Sudip Bhattacharyya$^{1}$,
Chandrachur Chakraborty$^{2}$,
A. R. Rao$^{1}$
\\
$^{1}$ Department of Astronomy and Astrophysics, Tata Institute of Fundamental Research, 1 Homi Bhabha Road, Mumbai 400005, India\\
$^{2}$ Kavli Institute for Astronomy and Astrophysics, Peking University, Beijing 100871, China
}

\date{Accepted XXX. Received YYY; in original form ZZZ}

\pubyear{2019}
\begin{document}
\label{firstpage}
\pagerange{\pageref{firstpage}--\pageref{lastpage}}
\maketitle

\begin{abstract}
An estimate of the jet inclination angle relative to the accreting black hole's spin can be useful to probe the jet triggering mechanism and the disc--jet coupling. A Tidal Disruption Event (TDE) of a star by a supermassive spinning black hole provides an excellent astrophysical laboratory to study the jet direction through the possibility of jet precession. In this work, we report a new method to constrain the jet inclination angle $\beta$ and apply it to the well-sampled jetted TDE Swift J1644+57. This method involves X-ray data analysis and comparisons of jet models with broad properties of the observed X-ray dips, to estimate the upper limit of the extent of the contribution of a plausible jet precession to these X-ray dips. From this limit, we find that $\beta$ is very likely to be less than $\sim 15^\circ$ for Swift J1644+57. Such a well-constrained jet inclination angle could be useful to probe the jet physics. The main advantage of our method is that it does not need to assume an origin of the observed X-ray dips, and the conclusion does not depend on any particular type of jet precession (e.g., the one due to the Lense-Thirring effect) or any specific value of precession frequency or any particular jet model. These make this method reliable and applicable to other jetted TDEs, as well as to other jetted accreting systems.
\end{abstract}




\section{Introduction}\label{intro}

Collimated jets are ubiquitous in accreting black hole systems, 
powered by accretion of matter in the vicinity of the black hole \citep{Blandford_Znajek_1977,Meier_2001}. But, as the environment near a black hole is not resolvable in high energies, important questions like the role of black hole spin (or its necessity) in jet triggering \citep{McKinney_2006} are yet to have any unanimously accepted answer. A related important problem of black hole astrophysics is whether a jet from an accreting black hole is aligned with the black hole spin axis or with the accretion disc angular momentum vector \citep{Fragile_2008mqw}, or with any other direction such as a magnetic axis, which is inclined with respect to the black hole spin axis \citep[e.g., ][]{Wang_2014}. An answer to this question could provide an important clue to the jet launching mechanism in black holes, disk-jet coupling, and the high energy radiation mechanisms in a black hole system \citep{Natarajan_1998, McKinney_2013}. 

If the jet is not aligned with the black hole spin, then it can precess and hence could produce a regular variation in the X-ray light curve, as the angle between the jet axis and the observer's line of sight changes during a precession period. But if the jet is aligned with the black hole spin axis, then it will not precess, and thus a corresponding variation in the X-ray light curve will not be observed.  These provide an observational tool to probe the jet direction.

Tidal Disruption Events (TDEs), or bursts of energy emission due to tidal disruption of stars by supermassive black holes (SMBHs) of typical masses of $10^{5-8}M_{\odot}$, are unique laboratories to probe various high energy astrophysical phenomena. 
Having a well defined and limited reserve of available accretable mass and hence a lifetime of only months to a few years, jetted TDEs are relatively clean systems to probe the jet alignment. In this paper, we model the X-ray data of a jetted TDE to constrain the jet direction.

Now, we briefly describe a few relevant points regarding TDEs. A star of mass $M_{*}$ and radius $R_{*}$ approaching an SMBH is subject to a strong tidal force exerted by the black hole. Once this force exceeds the self-gravity of the star (at the tidal radius, $R_{\rm t}$), the star is disrupted \citep{Rees_1988Nature}. 
The process is expected to happen up to $M_{\rm BH} \simeq 10^{8}M_{\odot}$ for a solar mass star.
Once disrupted, about half of the stellar material is ejected, and the remaining half becomes bound, returning to the pericentre and circularising, a fraction of which is accreted by the black hole \citep{Ayal_2000}. This produces a flare of electromagnetic radiation, usually peaked in the EUV or soft X-rays, which can reach up to Eddington or even super-Eddington luminosities \citep{Strubbe_Quataert_2009}. This emission is expected to fade with time as $t^{-p}$, with the index $p$ having different values depending on the model \citep{Guillochon_2013,Lodato_2011}.
 For example, $p$ may depend on the stellar composition \citep{Lodato_2009}. However, the most accepted canonical TDE luminosity decay trend is $t^{-5/3}$ \citep{Phinney_1989}. On top of this secular decay trend, the Lense-Thirring or any other type of precession may cause additional quasi-periodic dips. Note that among the sample of 87 TDE candidates till date (\url{http://tde.space/}), at least three are thought to harbour relativistic jets \citep{Auchettl_2017}. Among these jetted TDEs, Swift J164449.3+573451 (hereafter, J1644+57) is the most well known, well sampled and well studied one \citep{Burrows_nature_2011,Zauderer_nature_2011}.

\begin{figure}
\centerline{\includegraphics[width=1.0\linewidth]{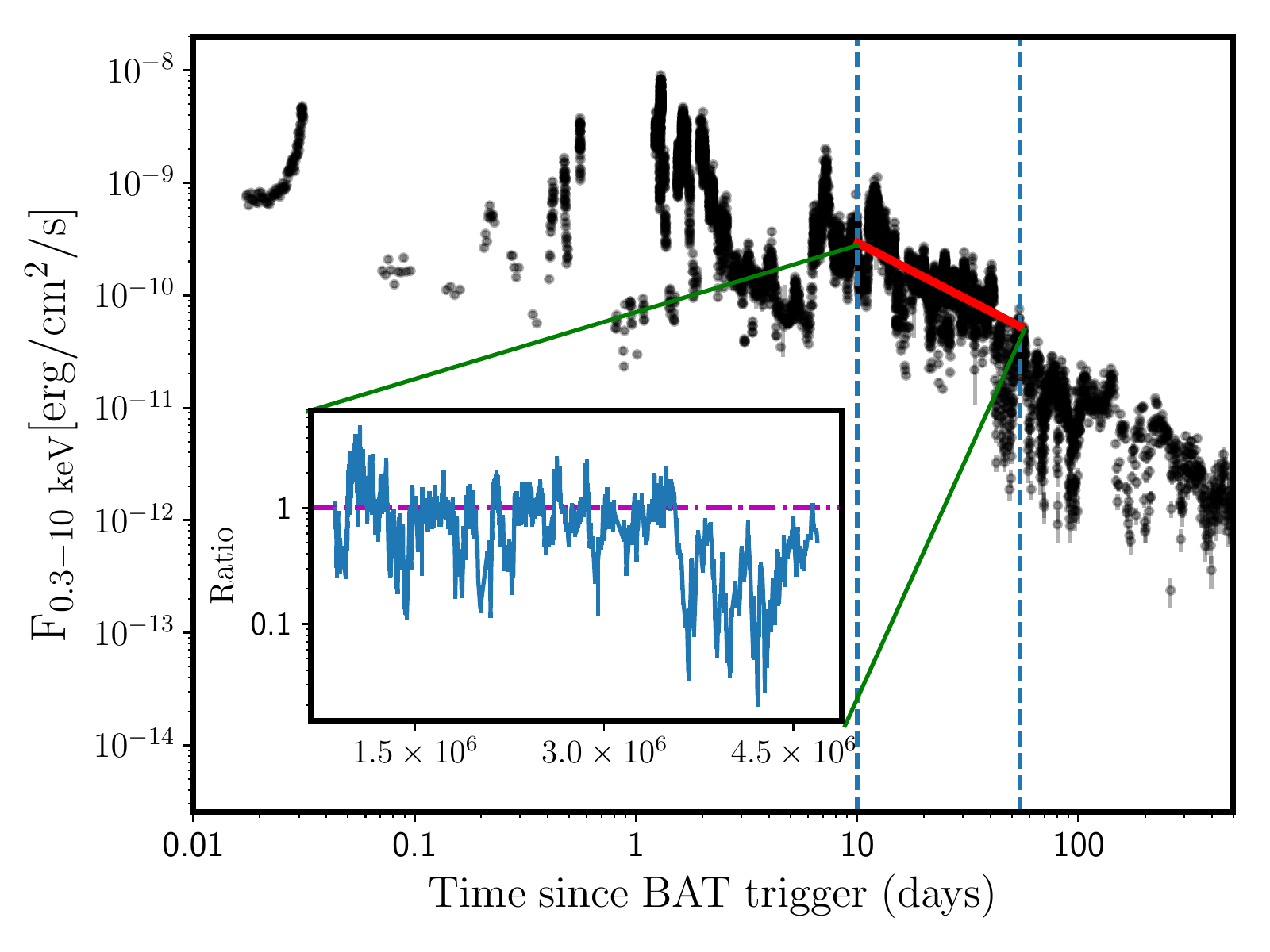}}
\caption{{\it Swift} XRT lightcurve of J1644+57: data taken from \citet{Mangano_2016} (which uses data from \url{http://www.swift.ac.uk/xrt\_curves/00450158/}), plotted along with a selected time range (which we investigate further) and the best-fit powerlaw continuum model (in red) in that time range. Inset: The data to model ratio from the selected time range in the main plot (section ~\ref{data_analysis}). The X-axis in the inset is in seconds (since {\it Swift}-BAT trigger )}.
\label{j1644_lc}
\end{figure}

\begin{figure}
\begin{subfigure}[b]{0.475\textwidth}
\centerline{\includegraphics[width=1.0 \linewidth]{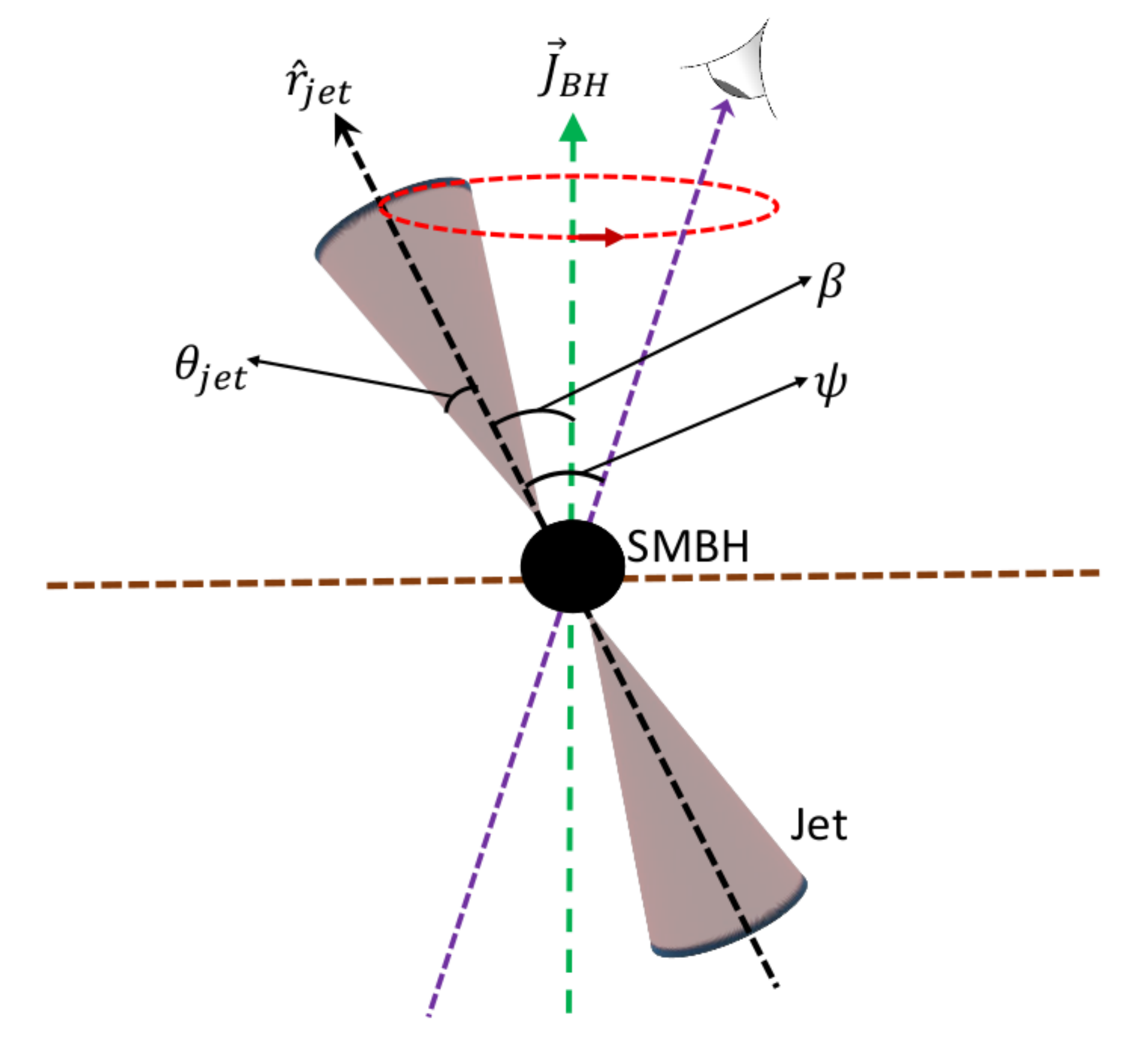}}
\end{subfigure}
\begin{subfigure}[b]{0.475\textwidth}
\centerline{\includegraphics[width=1.0 \linewidth, angle=0]{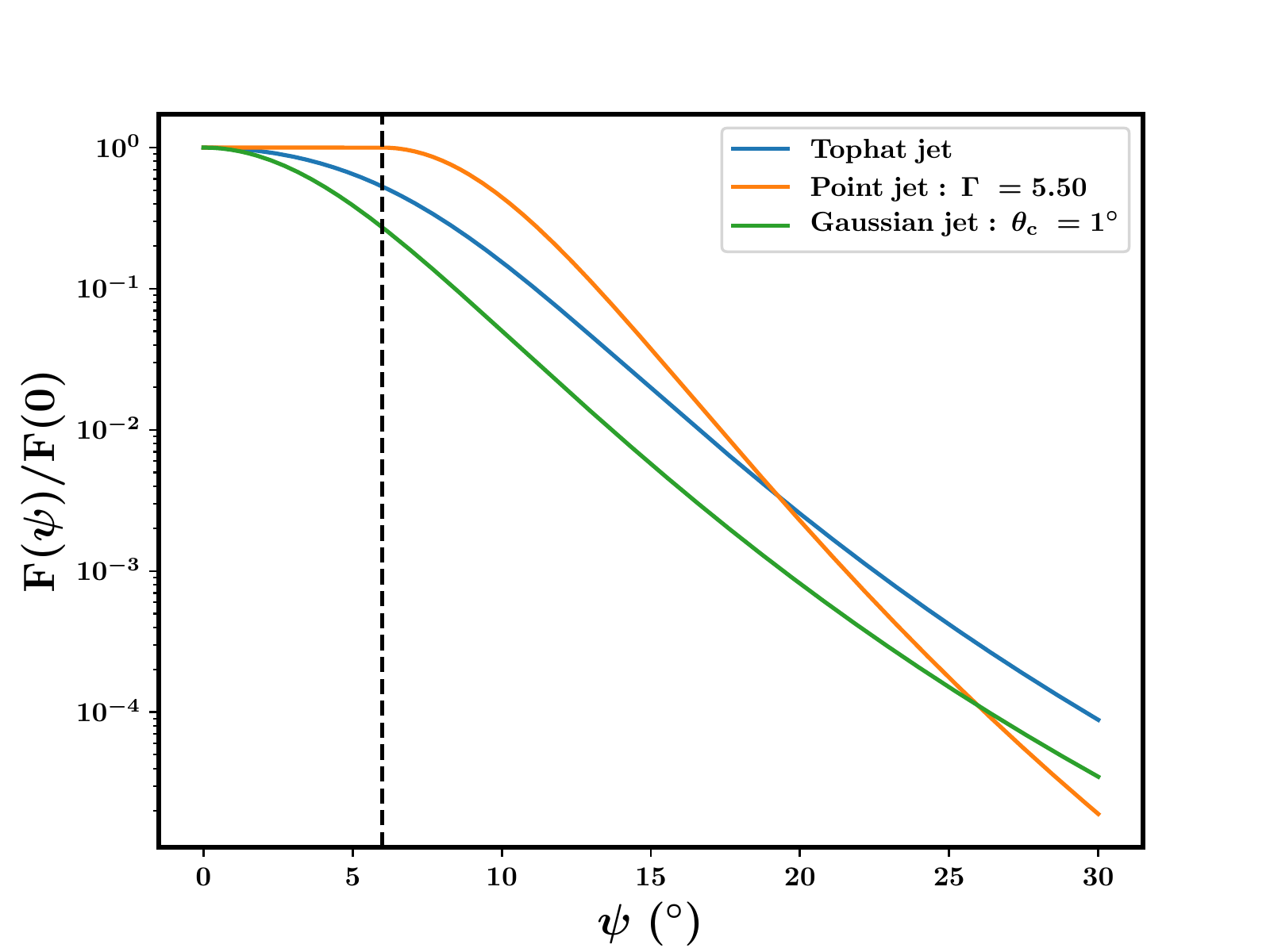}}
\end{subfigure}
\caption{Top panel: A schematic diagram representing our model of a jet from an accreting supermassive black hole. Bottom panel: Shows how the emission falls from the jet axis: comparison among  different jet models. The dashed vertical line marks the jet opening angle  $\theta_{\rm jet}$ (section~\ref{jet}).}
\label{lt_tde_model}
\end{figure}

J1644+57 was also the first observed jetted TDE, detected with {\it Swift} Burst Alert Telescope (BAT) on 28 March 2011 \citep{Burrows_nature_2011} in a star--forming host galaxy of redshift $z=0.3534$ \citep{Levan_2011}, and reaching X--ray luminosities $ > 10^{48} \rm{ \ erg \ s^{-1}} $. Detection of a  relativistic, highly collimated associated radio source \citep{Zauderer_nature_2011}, the rapid time variability in high energies, and a Blazar--like SED \citep{Burrows_nature_2011}, all indicated that J1644+57 had a compact relativistic jet directed almost towards our line of sight \citep{Bloom_2011}. The light curve of J1644+57 is rich in features \citep{Burrows_nature_2011, Mangano_2016}. The X-ray (0.3-10 keV) light curve started with a relatively flat trend for 10 days, accompanied by intense flaring (with a variability timescale of $\sim 100$s). Thereafter, the secular decay `roughly' followed the canonical $t^{-5/3}$ trend till $\sim 500$ days, after which there was a sudden drop in the X-ray emission ({\it Swift} light curve in Fig.~\ref{j1644_lc}). This drop is interpreted as the relativistic jet being switched off \citep{Zauderer_2013}. Comparing the observed rate of jetted TDEs with the theoretical rate, \citet{Burrows_nature_2011} reported (degenerate with each other) a jet opening angle (Fig.~\ref{lt_tde_model}) of $\theta_{\rm jet} \sim 5^{\circ}-6^{\circ}$ and a bulk Lorentz factor of $\Gamma \lesssim 20$. Moreover, the light curve of J1644+57 shows myriads of dips at different timescales, with QPO features reported at 200 s and 2.7 days. \citet{Saxton_2012} reported no change in the absorption column density $N_H$ during the dips, implying that they were not caused by absorbing clumps in the line-of-sight. The 2.7 days QPO feature, as well as its harmonics, have been proposed to be due to the precession of a misaligned jet \citep{Saxton_2012,Lei_2013}. On the other hand, the 200 s feature is thought to be associated with the corona or the Keplarian frequency at the ISCO \citep{Reis_2013}, or the inner jet in two component jet models \citep{Wang_2015}.

In this paper, we analyse the X-ray dips data from J1644+57, and probe the possibility of jet precession, without a priori assuming a particular model for these dips. In case there is a jet precession, we do not assume any specific periodicity of this precession, and do not try to explain the 2.7 days QPO. With our method involving minimum assumptions, we attempt to constrain the jet direction, and specifically to find a general upper limit to the jet inclination angle $\beta$ with respect to the black hole spin axis. In section~\ref{constraint}, we describe the data, and our models, methods and results. In section~\ref{Discussion}, we discuss our results.

\section{Constraint on jet inclination angle $\beta$}\label{constraint}

\subsection{Selection of X-ray data}\label{selection_data}

Inspired by the previous works, we impose the following two conditions to select the \textit{Swift} XRT data of J1644+57 dips (time range between two vertical lines in Fig.~\ref{j1644_lc}) for our analysis.

(a) Existence of a steady accretion flow: 
this happens when the circularization of the inward bound debris is complete, and a steady accretion flow sets in. Note that we expect the rigid precession of the disc (see next point) to kick in only after the circularization time ($t_{\rm circ}$; Eq. 7 in \citet{Ulmer_1999}). $t_{\rm circ}$ can be considered to be $\sim 10$ days, that is after the end of the flat, high flaring part of the X-ray lightcurve of J1644+57 (end of phase 0 in \citet{Shen_Matzner_2012,Shen_Matzner_2014}).

(b) Thick disc regime: 
 We consider the disc to be thick, if $H/R\gtrsim\alpha$, where $H$ is the disc thickness, $R$ is the radial distance from the SMBH centre and $\alpha$ is the dimensionless viscosity parameter in the range 0 to 1. 
In \citet[][hereafter SM12]{Shen_Matzner_2012} disc evolution scheme, this corresponds to phase 1. In this regime (from $t_{\rm circ}$ till $t_{\rm thin}$; \citet{SL_2012}), the disc is geometrically thick and super-Eddington \citep{Mangalam_TDE_2018}, and the fallback rate could follow the canonical $t^{-5/3}$. Assuming the form of $t_{\rm thin}$ from Eq. 6 of \citet{SL_2012} \citep[which assumes][disk surface density profile]{Strubbe_Quataert_2009}, one can estimate the $t_{\rm thin}$ value for realistic values of parameters. Taking the acceptable ranges of $M_{\rm BH}, M_{*}$ and $\alpha$ \citep{Burrows_nature_2011,Zauderer_nature_2011,Tchekhovskoy_2014}, assuming a typical mass-radius relation for main sequence stars \citep{Demircan-Kahraman_1991}, noting that the typical values of $\alpha$ should be much smaller than 0.1 for the disk precession in a TDE disk \citep{Nixon_2013}, and considering the dips and previous studies of the X-ray light curve (e.g., \citet{Saxton_2012}), we consider $t_{\rm thin}$ to be $\sim4.7\times10^{6}$ s or $\sim$54 days.
In a TDE, the star will not generally approach the black hole in its spin-equatorial plane, and hence the initially formed thick accretion disc is expected to be tilted \citep{Fragile_2007} with respect to the black hole spin. Note that in the later thin disk regime of a TDE, such a tilt of the inner disk might cease to exist due to the Bardeen-Petterson effect  (\citet{Bardeen_Petterson_1975}; but see \citet{Chakraborty_Bhattacharyya_2017}). A tilted thick disc would precess by the Lense-Thirring (LT) effect \citep{Bardeen_Petterson_1975}, and if the jet is attached to the disc, it could also precess with such a disc. Therefore, while in our study we consider any general tilt (with respect to the black hole spin) and precession of the jet, we consider only the thick disc regime, because a particular precession, viz., the LT precession, of the jet may happen only in this regime.

Therefore, we use 44 days of  \textit{Swift} XRT data (from 10 days to 54 days since the BAT trigger) for our analysis.

\subsection{Analysis of X-ray data}\label{data_analysis}

In order to identify the dips from the \textit{Swift} XRT data of J1644+57, we first determine $p$ of the X-ray intensity time evolution function $t^{-p}$ (see section~\ref{intro} and Fig.~\ref{j1644_lc}). Firstly, we adopt the spectra-dependent, ECF corrected unabsorbed XRT flux lightcurve from \citet{Mangano_2016}. Thereafter, we select the mentioned time window from $t_{\rm circ}$ till $t_{\rm thin}$ (Fig.~\ref{j1644_lc}; section~\ref{selection_data}). In order to get the secular trend, the selected data are smoothed (with boxcar) at different timescales, fitted with a $t^{-p}$ powerlaw, and the average of the best-fit $p$ values over all such timescales is taken. This average best-fit $p$ comes out to be $0.79 \pm 0.01$. This is consistent with \citet{Mangano_2016}, but is inconsistent with the canonical value of $5/3$. Subsequently, we calculate the ratio between the data and the best-fit continuum, and identify dips following the method proposed by \citet{Mangano_2016}. These identified dips in the ratio curve are used to investigate the jet inclination angle $\beta$.

\subsection{Salient points of our method to constrain $\beta$}\label{salient}

Here, we summarize the salient aspects of our method. The method has been described in more detail in section~\ref{our_method}.

The origin of the dips in the X-ray light curve of J1644+57 is not known. Since the X-rays in this phase are primarily from the jet (see section~\ref{intro}), the jet precession, by which the jet periodically and partially can go out of the sight, might have caused these dips. But it is clear from the data that diverse, irregular dips of the X-ray light curve (Fig.~\ref{j1644_lc}) cannot  match well with the regular modulations expected from the pure jet precession. So, one extreme assumption that X-ray dips were purely due to jet precession, cannot be correct. It has also been proposed that none of the dips were due to jet precession \citep{SL_2012}. However, there is no guarantee that this assumption (which is the other extreme) is correct either. The complex dips could originate due to more than one physical process, such as jet precession, intrinsic jet intensity variations due to fluctuations of physical quantities, instabilities, effect of winds, episodic jet activities and so on. In our method, unlike previous works, we consider this general and realistic scenario. We neither assume that dips were not at all due to jet precession \citep[like in ][which could be an unrealistic assumption]{SL_2012}, nor assume that dips were entirely due to jet precession. We consider that jet precession could contribute to dips.

If the jet precesses,  it is expected to be inclined with respect to the black hole spin axis. The amount of this inclination, i.e., the angle $\beta$, would affect the extent of contribution of jet precession to the X-ray dips. Therefore, an estimation of the extent of this  contribution would enable us to constrain the jet inclination angle $\beta$. In order to achieve this, we first construct a theoretical jet precession model (for three types of jet; section~\ref{jet}), and the corresponding light curve. Then we identify two observed parameters, vis., fractional amplitude and ratio (defined and described in sections~\ref{amplitude} and \ref{ratio}) related to X-ray dips, and impose the minimum condition for them to be due to the jet precession. From this bare minimum requirement, we estimate the upper limit of the extent of contribution of jet precession to X-ray dips. From this upper limit, we find an upper limit of $\beta$ using our jet models (see section~\ref{our_method}).

\subsection{Theoretical jet models}\label{jet}

We use the following three reasonably different jet models to show the robustness of our results.

(a) Point--like jet: Here, similar to \citet{Lei_2013}, we assume that the jet emission is constant within the opening angle $\theta_{\rm jet}$, and falls off outside the jet (Fig.~\ref{lt_tde_model}). The flux density at a frequency $\nu$ and time $t$, for an observation angle (measured from the jet axis; see Fig.~\ref{lt_tde_model}) $\psi (> \theta_{\rm jet}$), is given as \citep{Granot_2002}:
\begin{equation}
\label{flux}
F_{\rm \nu}(\psi, t)=D^{3}F_{\rm \nu/D}(0,Dt)
\end{equation}
where $D=\frac{1-B}{1-B\cos(\psi)}$ and $B=\sqrt{1-\frac{1}{\Gamma^{2}}}$. 
Assuming a $t^{-p}$ secular trend of the light curve and a $\nu^{-\alpha}$ spectral shape (where we have used two values of $\alpha$, motivated by \citet{Saxton_2012}:  1.7 within $\theta_{jet}$ and 2.8 outside $\theta_{jet}$ ), the X-ray flux ($ F(\psi,t)=\nu F_{\rm \nu}(\psi, t)$) becomes:
\begin{equation}
\label{SED}
F\left(\psi,t\right)=D^{4-p+\alpha}F\left(0,t\right)
\end{equation}
The two values of $\alpha$ (1.7 and 2.8) are assumed to be the spectral index values within the jet core and outer part, respectively. This is motivated by the flux-resolved energy spectra from \citet{Saxton_2012}, where a change of spectral index from 1.7 to 2.8 could be observed as a function of the depth of the dips.

(b) Top--hat jet: In this case and in the following one, we assume a homogeneous, geometrically and optically thin jet \citep{salafia_2015}. Then:
\begin{equation}
\label{salafia_flux}
\frac{F\left(\psi\right)}{F(0)}=\frac{\int_{0}^{2\pi}\int_{0}^{\pi/2}\frac{\epsilon(\theta)\sin\theta d\theta d\phi}{\left[1-B\left(\cos\theta\cos\psi+\sin\theta\sin\phi\sin\psi\right)\right]^{3+\alpha\left(\theta\right)}}}{\int_{0}^{2\pi}\int_{0}^{\pi/2}\frac{\epsilon(\theta)\sin\theta d\theta d\phi}{\left[1-B\cos\theta\right]^{3+\alpha\left(0\right)}}}
\end{equation}
where $\theta$ and $\phi$ are the spherical angles in the jet frame, and $\epsilon(\theta)$ is the energy emitted by the part of the jet comprised between $\theta$ and $\theta+d\theta$, divided by the corresponding solid angle. $\epsilon(\theta)$ is thus a quantification of the intrinsic jet structure. Again, the spectral index $\alpha$ is defined for within the jet core and outer part as follows (motivated by \citet{Saxton_2012}):
\begin{equation}
\label{alpha}
\alpha(\theta)=
\begin{cases}
    1.7,& \text{for } \theta \le \theta_{jet}\\
    2.8,              & \text{for } \theta > \theta_{jet}
\end{cases}
\end{equation}
For the top--hat jet, $\epsilon(\theta)$ (normalized to 1) can be taken as:
\begin{equation}
\label{tophat}
\epsilon(\theta)=
\begin{cases}
    1,& \text{for } \theta \le \theta_{jet}\\
    0,              & \text{for } \theta > \theta_{jet}
\end{cases}
\end{equation}
Plugging equation \ref{tophat} into equation \ref{salafia_flux}, we get the off--axis flux for a top--hat jet as a function of viewing angle $\psi$.

(c) Gaussian jet: Here, the intrinsic structure can be taken as \citep{Kumar_granot_2003}:
\begin{equation}
\label{gaussian}
\epsilon\left(\theta\right) \propto e^{-\left(\theta/\theta_{c}\right)^{2}}
\end{equation}
where we have assumed $\theta_c$ to be $\sim 1^\circ$.

A comparison between off--axis emissions from these three jets is given in Fig.~\ref{lt_tde_model} (lower panel).

\subsection{Our method to constrain $\beta$}\label{our_method}

As indicated in section~\ref{salient}, we use the following two new methods to estimate two upper limits of the jet inclination angle $\beta$ for each jet model. We then combine these two limits, separately for each type of jets, to obtain an upper limit of $\beta$ (Fig.~\ref{beta_a_us}).

\subsubsection{Fractional amplitude method}\label{amplitude}

If the jet is inclined and precesses, then a regular variation in the observed X-rays is expected (sections~\ref{intro} and \ref{salient}). For a point--like jet, such a modulation will happen only if the observation angle $\psi$ is greater than $\theta_{\rm jet}$ during a part of the precession period (section~\ref{jet}; Fig.~\ref{lt_tde_model}). The maximum fractional depth of this modulation, viz., the fractional amplitude $f_{\rm d}$, can be expressed using our jet model formulae (section~\ref{jet}) in the following definition:
\begin{equation}
\label{Off-axis_trend}
f_{\rm d} = (F(0)-F(\psi_{\rm max}))/F(0).
\end{equation}
Here, $\psi_{\rm max}$ is the maximum $\psi$ value in a jet precession period.
As expected (and as found from our theoretical model) $f_{\rm d}$ increases as $\beta$ and $\Gamma$ increase. So, an upper limit of $\beta$ can be obtained from an observationally inferred upper limit of $f_{\rm d}$, when compared with the model values for a lower limit of $\Gamma$. The value of $\Gamma$, as reported at very early stages of this TDE, is ~10 \citep{Bloom_2011,Burrows_nature_2011,Lei_2013}. The $\Gamma$ could decrease after this (similar to the jets in GRB afterglow). So we assume that $\Gamma$=5.5 (following \citet{Wang_2014} and consistent with \citet{Lu_2017}). While the limit on the jet opening angle ($\theta_{\rm jet}$) is uncertain, all of the available literature point towards $\theta_{\rm jet}$ value of 5-6$^\circ$ \citep{Bloom_2011,Burrows_nature_2011,Lei_2013}. We have thus chosen a reasonable value of $\theta_{\rm jet}= 6^\circ$ for further analysis.

We consider a very general scenario that the X-ray dips could have several origins at a time (including jet precession (see section~\ref{salient})) and hence several physical effects are possibly mixed in the dips. Here we describe our method to estimate the upper limit of $f_{\rm d}$ (which is purely due to jet precession) from the observed X-ray light curve. Let us first assume that the jet precession period, as seen by the observer, is $T_{\rm box}$.
Note that the intrinsic period $T_{\rm prec}$ is related to $T_{\rm box}$ by $T_{\rm box} = (1+z) \times T_{\rm prec}$, where $z$ is the redshift. Then the observed X-ray modulation (suppose, with the fractional amplitude $f_{\rm d}$) due to the jet precession \textit{alone} should repeat in every time box of width $T_{\rm box}$ in our data set. Hence a dip with the fractional amplitude of at least $f_{\rm d}$ should be present in each of these time boxes, and this is a minimum condition (indicated in section~\ref{salient}) for a `jet precession induced modulation' with the fractional amplitude $f_{\rm d}$ to be present. Therefore, we run a sliding time box of width $T_{\rm box}$ in the time window from $t_{\rm circ}$ till $t_{\rm thin}$ in the X-ray data (Fig.~\ref{j1644_lc}). We collect the fractional depth ($f_{\rm d}^{\rm obs}$) of the deepest dip in each such box. Then we choose the minimum value ($f_{\rm d}^{\rm obs, min}$) from all the $f_{\rm d}^{\rm obs}$ values. This means that every time box of width $T_{\rm box}$ has a dip with the fractional amplitude of at least $f_{\rm d}^{\rm obs, min}$, and hence the jet precession could cause an X-ray intensity modulation with the fractional amplitude $f_{\rm d}^{\rm obs, min}$. However, other physical effects (see section~\ref{salient}) could also contribute to this fractional amplitude of $f_{\rm d}^{\rm obs, min}$ partially or fully, and hence the measured $f_{\rm d}^{\rm obs, min}$ is the upper limit of $f_{\rm d}$. As mentioned in the previous paragraph, we compare our model values for each type of jets with this upper limit to estimate an upper limit of $\beta$.

\begin{figure}
\begin{subfigure}[b]{0.45\textwidth}
\centerline{\includegraphics[width=0.9 \linewidth]{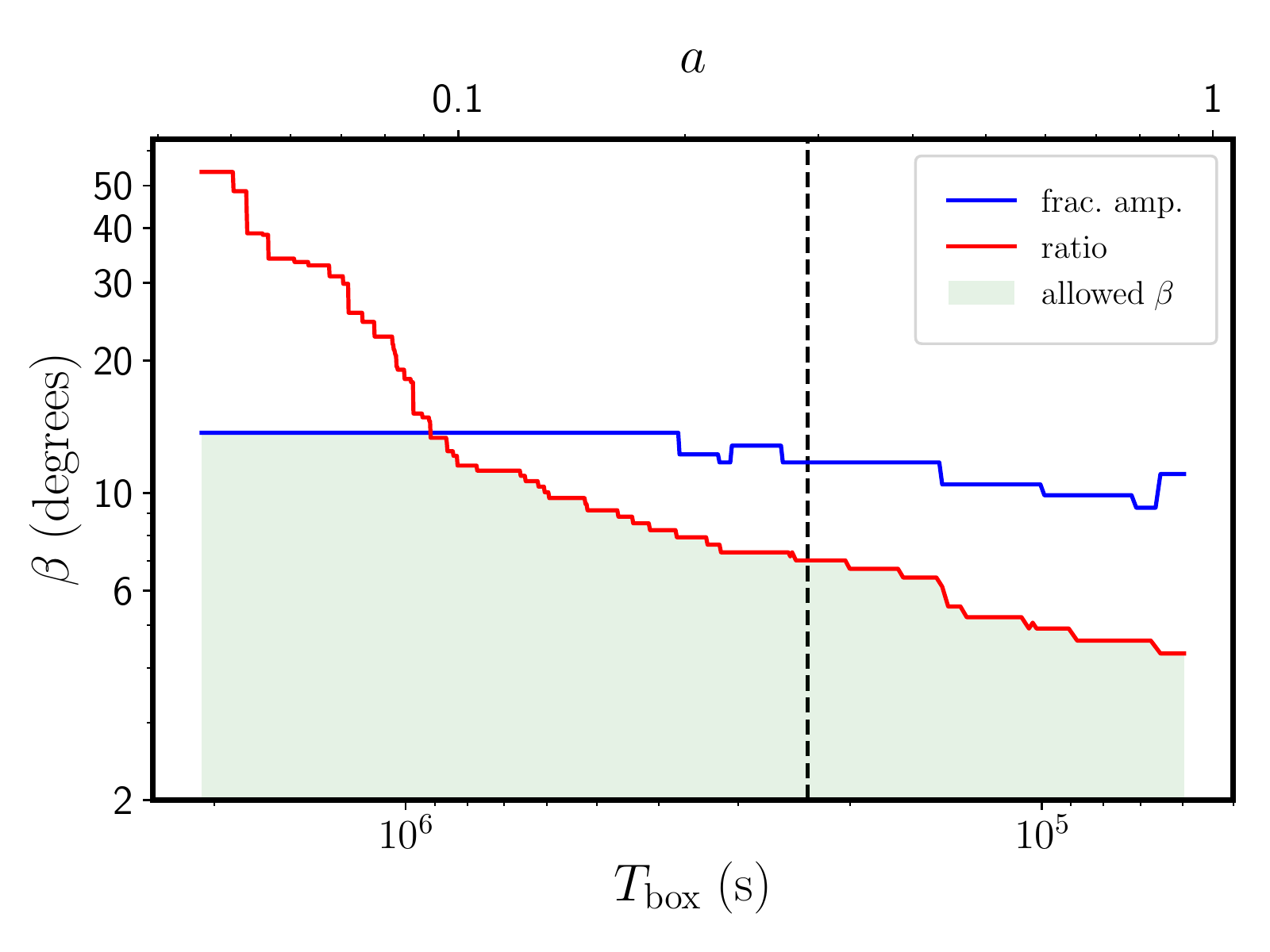}}
\end{subfigure}
\begin{subfigure}[b]{0.45\textwidth}
\centerline{\includegraphics[width=0.9 \linewidth, angle=0]{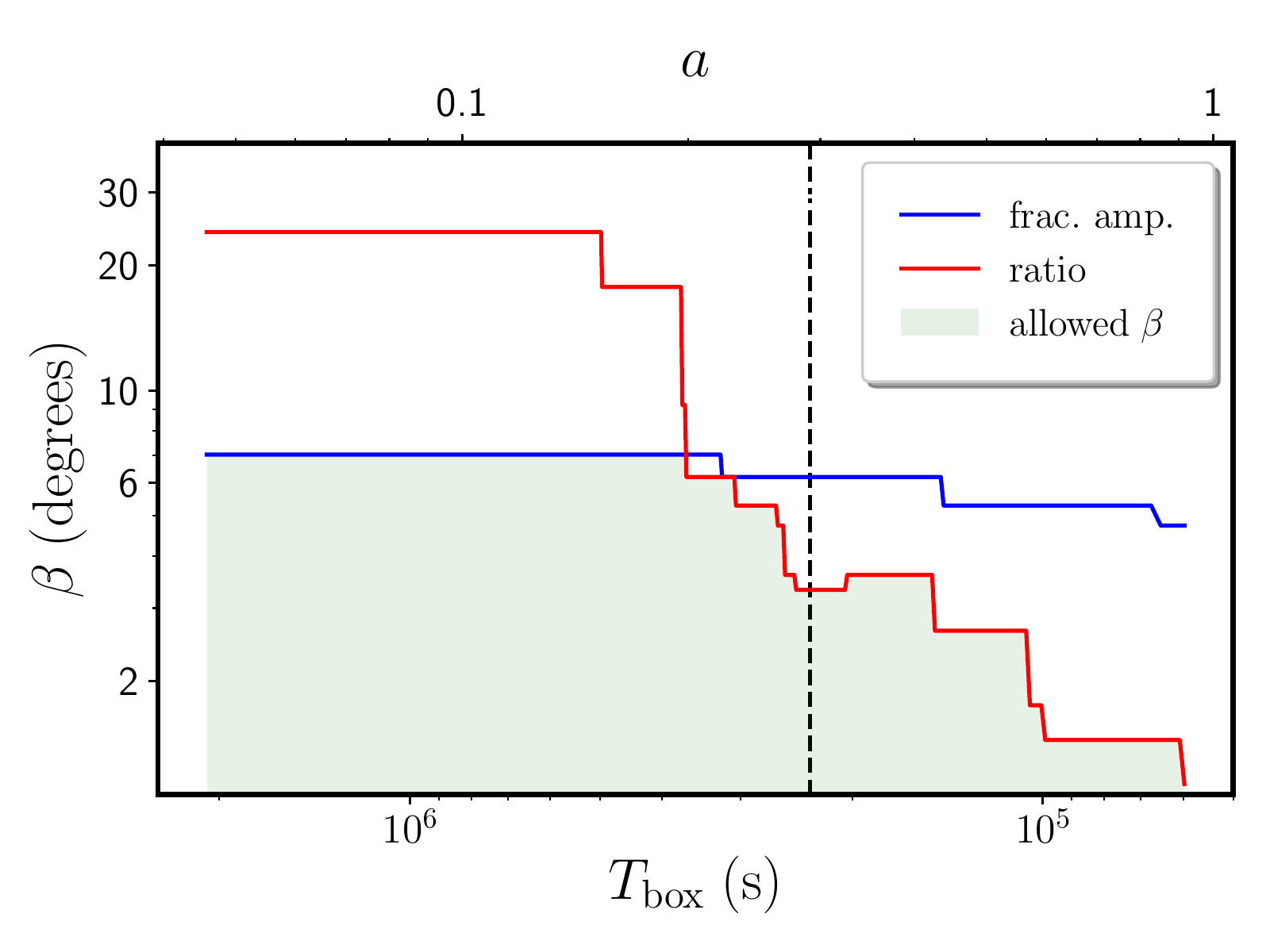}}
\end{subfigure}
\begin{subfigure}[b]{0.45\textwidth}
\centerline{\includegraphics[width=0.9 \linewidth, angle=0]{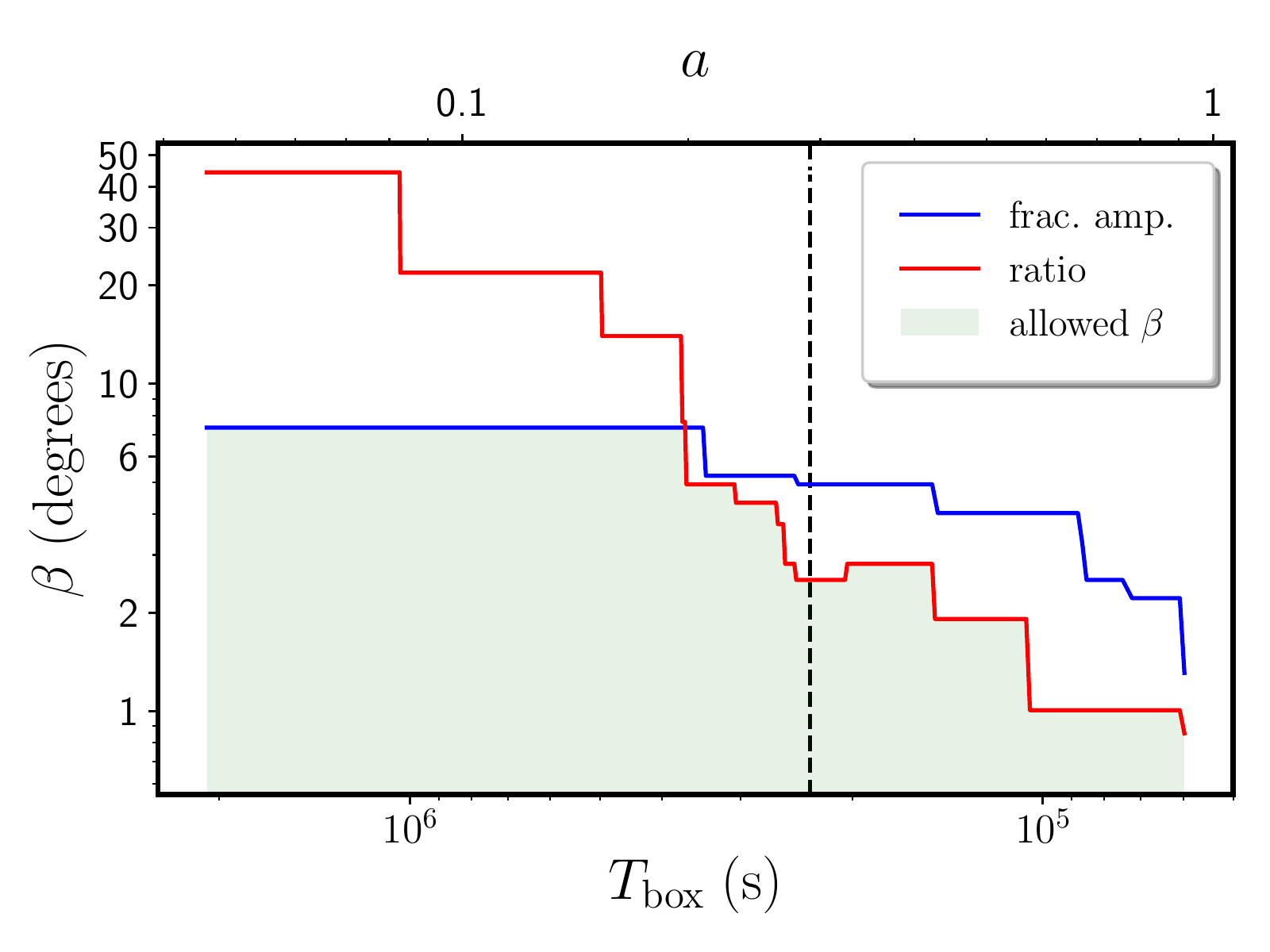}}
\end{subfigure}
\caption{A panel shows our results for a particular type of jet: top panel: point--like jet; middle panel: top--hat jet; and lower panel: Gaussian jet (section~\ref{jet}). Each panel shows our estimate of the upper limit of the jet inclination angle $\beta$ as a function of the jet precession period $T_{\rm box}$ (see section~\ref{our_method}). The upper x-axis shows the black hole spin parameter, only if the jet precession is due to the Lense-Thirring effect. The blue and red curves are upper limits of $\beta$ from the fractional amplitude method (see section~\ref{amplitude}) and the ratio method (see section~\ref{ratio}), respectively. The common area under these two curves, where both the upper limits are simultaneously valid, gives the allowed values of $\beta$. It is evident from the plots for all jet types that the value of $\beta$ is low, and hence the jet axis is close to the black hole spin axis. The vertical dashed black line corresponds to a plausible 2.7 days jet precession period (guessed from a QPO observed with this period; see section~\ref{intro}).}
\label{beta_a_us}
\end{figure}

This upper limit of $\beta$ is for an assumed jet precession period $T_{\rm box}$. Since we do not know the jet precession period (in case the jet precesses), we repeat the above exercise for many $T_{\rm box}$ values from the range $5\times10^{4}$s to $2.5\times10^{6}$s to make our conclusion robust.
Here, the minimum box size was taken so to ensure at least 10 data points in each of the boxes (for sufficient statistics), and the maximum box size was selected so that we have sufficient number of sliding boxes in our data set. Therefore, we get the upper limit of $\beta$ as a function of $T_{\rm box}$ (blue curves in three panels of Fig.~\ref{beta_a_us} for three types of jet).

\subsubsection{Ratio method}\label{ratio}

In case of jet inclination and precession, in addition to $f_{\rm d}$, we define another parameter, ratio ($R$), in the following way. First, we set a reference fractional flux level at $k = F/F(0)$.
Then we define $R = N_{\rm above}/N_{\rm below}$, where $N_{\rm above}$ is the number of fractional flux points above $k$,
while $N_{\rm below}$ is the number of those below $k$. 
As expected, and as found from our model calculations, $R$ decreases as $\beta$ and $\Gamma$ increase. So, 
an upper limit of $\beta$ can be obtained from an observationally inferred lower limit of $R$, when compared with the model values for a lower limit of $\Gamma$ (see section~\ref{amplitude} for $\theta_{\rm jet}$ and $\Gamma$ limit values).

Similar to the $f_{\rm d}$ method (section~\ref{amplitude}), it is challenging to observationally estimate the lower limit of $R$, because of plausible multiple physical origins of X-ray dips at a time (section~\ref{salient}). To achieve this, we use the similar general procedure mentioned in section~\ref{amplitude}, as follows.
For a specific $T_{\rm box}$ value, we collect the observationally estimated $N_{\rm above}/N_{\rm below}$ values (i.e., $R^{\rm obs}$) from sliding boxes. For this, we use $k=0.73$, so that for all time boxes mentioned in section~\ref{amplitude}, $k > F(\psi_{\rm max})/F(0)$ (see Eq.~\ref{Off-axis_trend}). If the modulation of the observed X-ray light curve were purely due to jet precession, then the $R^{\rm obs}$ values from all sliding boxes would be the same (i.e., $R$) within the statistical errors. However, the plausible additional contributions to the dips due to other physical effects (see above) would make the observational ratio $N_{\rm above}/N_{\rm below}$ lower than $R$, and hence $R^{\rm obs} \le R$. This is a minimum condition (indicated in section~\ref{salient}) for a `jet precession induced modulation' with the ratio $R$ to be present. Since, the maximum ($R^{\rm obs,max}$) of the $R^{\rm obs}$ values from all time boxes is the least affected by other physical effects, $R^{\rm obs,max}$ is the observationally inferred lower limit of $R$. As mentioned in the previous paragraph, we compare our model values for each type of jets with this lower limit to estimate an upper limit of $\beta$. Repeating the same exercise for many $T_{\rm box}$ values, we get $\beta$ as a function of $T_{\rm box}$ (red curves in three panels of Fig.~\ref{beta_a_us} for three types of jet).

Note that the upper limit of $\beta$ increases (see both blue and red curves of Fig.~\ref{beta_a_us}) for increasing $T_{\rm box}$, because wider time boxes should have 
more larger dips, thus increasing the upper limit of the observed fractional amplitude and decreasing the lower limit of the observed ratio. Applying the $f_{\rm d}$ method and the $R$ method jointly, we get a general upper limit of $\beta$ (the upper limit of the common area under the two curves in each panel of Fig.~\ref{beta_a_us}).

\subsubsection{Special case: LT precession }\label{special}

The results of sections~\ref{amplitude} and \ref{ratio} are for any general direction of the jet and any general jet precession frequency. Let us now consider a special case, where the jet is attached to a tilted accretion disc, and hence is precessing with the LT frequency. In this case, $T_{\rm box} = T_{\rm prec}^{\rm LT}$, where $T_{\rm prec}^{\rm LT}$ is the LT precession time period.

As opposed to the simple and commonly used expression of $T_{\rm prec}^{\rm LT}$ \citep[e.g., in ][]{SL_2012}, which is valid only for very small values of the dimensionless black hole Kerr or spin parameter $a$, here we use the fully general relativistic (GR) expression. This is calculated by putting the full GR expressions of related quantities from \citet{Kato_1990, Chakraborty_Bhattacharyya_2017, Chakraborty_Kocherlakota_2017} into the $T_{\rm prec}^{\rm LT}$ defined in \citet{Liu_Melia_2002}. If $\tau$ is the torque on the precessing tilted accretion disc (tilted at an angle $\beta$) and $J$ is its angular momentum \citep{Liu_Melia_2002,Fragile_2007}, the Lense-Thirring precession time period is given by 
\begin{equation}
\label{T_prec}
T_{\rm prec}^{\rm LT}=2\pi\sin\beta\frac{J}{\tau}.
\end{equation}
The $T_{\rm prec}^{\rm LT}$, thus calculated, depends on the following parameters.
(1) Surface density profile: we use the phenomenological surface density profile $\Sigma=\Sigma_{\rm i}\left(r/r_{\rm i}\right)^{\rm -\zeta}$ with $\zeta = 1$. 
(2) Inner and outer edge of accretion disc: we fix the inner radius ($r_{\rm i}$) of the accretion disc at the innermost stable circular orbit (ISCO) of the Kerr black hole \citep{Bardeen_1972}, and the outer radius ($r_{\rm o}$) at the tidal disruption radius ($R_{\rm T}$) or twice the pericentre distance ($R_{\rm p}$) \citep{SL_2012,Tchekhovskoy_2014}.
(3) Mass ($M_{\rm BH}$) and Kerr parameter ($a$) of the SMBH: we use $M_{\rm BH} = 10^6 M_{\odot}$, and thus we can have a one-to-one correspondence between $T_{\rm box}$ and $a$.
The full expression of $T_{\rm prec}^{\rm LT}$ cannot be expressed analytically, and is therefore numerically computed.

Using the the relation between $T_{\rm box}$ and $a$, we express the upper limit of $\beta$ as a function of $a$ (see the upper x-axis of Fig.~\ref{beta_a_us}).

\subsection{A method assuming the jet precession does not contribute to dips}\label{optimistic}

\begin{figure}
\centerline{\includegraphics[width=0.8 \linewidth]{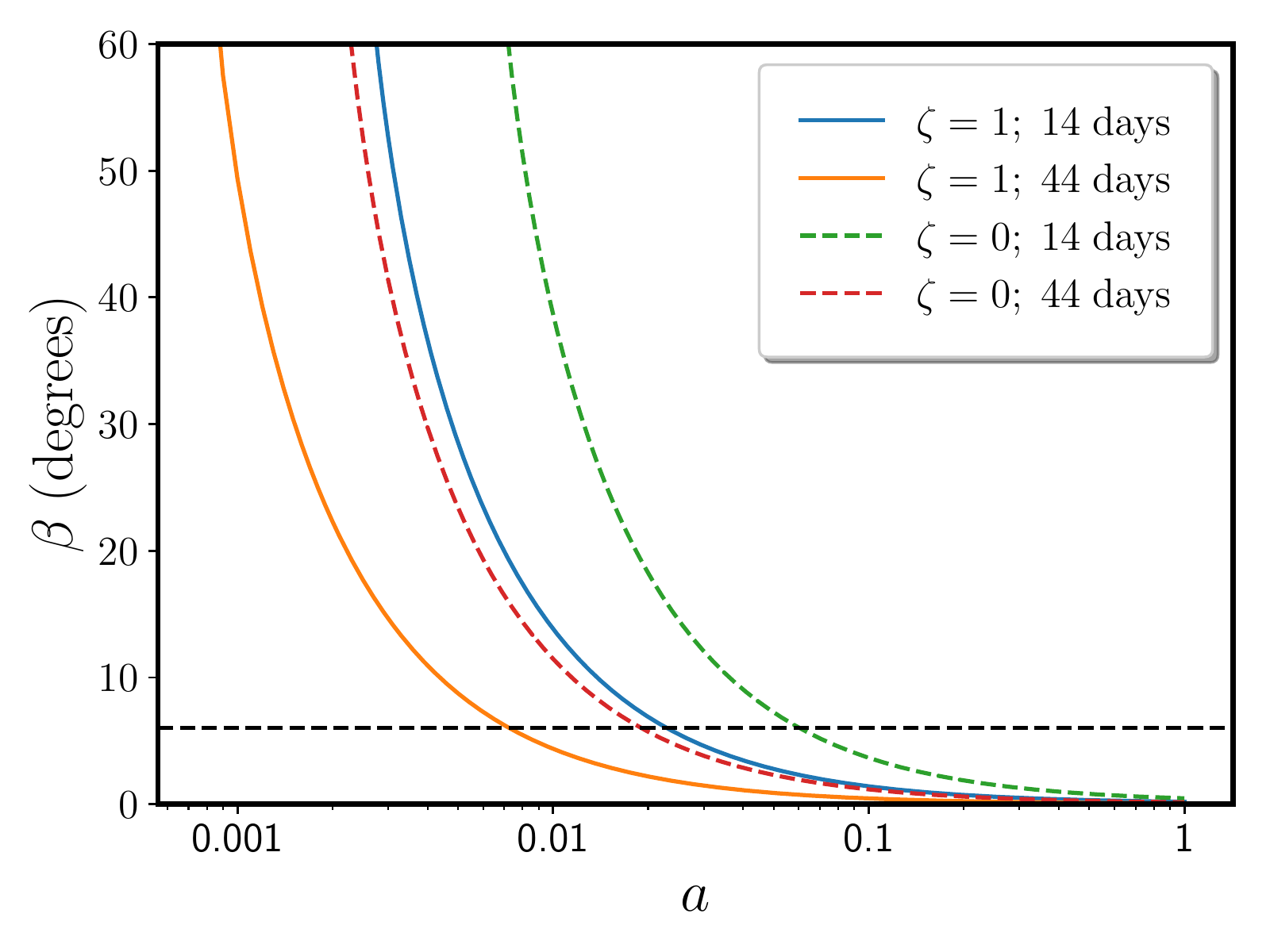}}
\caption{Upper limit of the jet inclination angle $\beta$ as a function of the black hole Kerr parameter $a$ for J1644+57.
This method assumed that none of the observed dips in the {\it Swift} XRT light curve of J1644+57 were due to the precession of the jet in 14 days \citep{SL_2012} and in 44 days (this work; see section~\ref{optimistic}). For each curve, the parameter space to the left of the curve is allowed. It can be seen that the more extended data coverage of this paper gives more stringent constraints on $\beta$ for a given $a$ (see section~\ref{optimistic}). The black horizontal line corresponds to the jet opening angle of $6^{\circ}$, and this plot uses the LT precession time period formula discussed in section~\ref{special}. However, this plot requires several assumptions \citep[following ][]{SL_2012}, which may not be entirely reasonable (section~\ref{optimistic}).
 }
\label{beta_a_SL}
\end{figure}

If we make the following assumptions (similar to \citet{SL_2012}):\\
(1) The jet precession does not contribute to X-ray dips observed from J1644+57 in 44 days (between $t_{\rm circ} \sim $10 days and $t_{\rm thin} \sim$ 54 days).\\
(2) The jet is inclined with respect to the black hole spin axis, it is aligned with the accretion disc angular momentum vector, and precesses with the LT precession frequency of the disc.\\
(3) The jet is like our point-like jet (section~\ref{jet}), for which the emission intensity is constant within the jet opening angle.\\
 Then we can get an optimistic constraint on $\beta$ (Fig. \ref{beta_a_SL}).

 Note that \citet{SL_2012} assumed that the precession of the point-like jet did not contribute to X-ray dips observed from J1644+57 in 14 days. This choice was probably made due to the unavailability of long-term data at that time. As we argued earlier (section~\ref{salient}), we cannot rule out such a contribution at this time. We, on the other hand, have not made any of these assumptions (sections~\ref{amplitude} and \ref{ratio}). We have modelled the X-ray data and considered three different types of jets to estimate an upper limit of $\beta$. But if we do make such assumptions, we can put a much more stringent upper limit on $\beta$, as we consider more X-ray data (44 days instead of 14 days; see Fig.~\ref{beta_a_SL}).

\vspace{0.4cm}
\section{Discussion}\label{Discussion}

In this paper, we report a new method to estimate the upper limit of the jet inclination angle $\beta$ with respect to the black hole spin axis. We apply this method to the jetted TDE J1644+57. The main advantage of this method is that one does not have to make extreme, and plausibly unrealistic, assumptions, such as (1) the contribution of a likely jet precession to observed X-ray dips is nil; or (2) a jet precession is the sole cause of X-ray dips. Rather, our method considers a very general scenario that a jet precession could somewhat contribute to the observed X-ray dips. So we do not need to assume an origin of the observed X-ray dips. We simply estimate the upper limit on the extent of the contribution of jet precession to these dips, by imposing two minimum conditions on two parameters of the modulation due to jet precession: fractional amplitude (section~\ref{amplitude}) and ratio (section~\ref{ratio}). Note that,
while the first parameter serves as a proxy for the depth of the modulation, the second parameter serves as a proxy for the shape of this modulation. From the above mentioned limit on extent, we find that $\beta$ is very likely to be less than $\sim 15^{\circ}$ (Fig.~\ref{beta_a_us}). This conclusion does not depend on any particular type of precession (e.g., LT precession) and any specific value of precession frequency, and also remains same for different types of jets. All these make our method more robust (e.g., relative to that of \cite{SL_2012}; see section~\ref{optimistic}), and general enough for application to other similar systems, such as jetted TDE lightcurves for different wavelengths of observation, and potentially for non-jetted TDEs. Note that, even though we are unable to estimate a $\beta$ upper limit for $T_{\rm box} > 2.5\times10^{6}$~s, an upper limit of $\sim 15^{\circ}$ is indicated by the extrapolation of the almost horizontal blue curves in Fig.~\ref{beta_a_us}.

Some of the parameters in the jet models used in this work are somewhat poorly constrained. However, the choices of the parameter values are backed up by the available literature on J1644+57. The main plausible sources of uncertainties and how they were dealt with, are as follows. In this work, the errors on the flux values in the X-ray light curve were not considered, as they are typically very small. We further investigated two additional sets of ratio between the data and the best-fit $t^{-p}$ model of the light curve: the upper and lower edges of the error envelope of the detrended light curve. The resultant change was found to be very small. We have also checked that the small uncertainty in powerlaw index ($p$) in the $t^{-p}$ model of the light curve, affects the final result negligibly. Hence, we neglect this uncertainty. Additionally, we have used two values of the spectral index $\alpha$ in all the jet models, as seen in eq.~\ref{alpha} of section~\ref{jet}. In order to check how the $\alpha$ values affect our results, instead of assuming a step function, we have also tested with two extreme values of $\alpha$: $\alpha$=1.7 throughout, and $\alpha$=2.8 throughout. Comparing the results from either of these two cases with with our original jet models (section~\ref{jet}), we found that the differences are negligibly small. Furthermore, although the values of the jet opening angle ($\theta_{jet}$) and bulk Lorentz factor ($\Gamma$) are somewhat uncertain and can induce systematic uncertainty in the result, the justification of the choices of these values have been elaborated in section~\ref{amplitude}. Finally, for the Gaussian jet, the characteristic angle $\theta_c$ is not a directly measurable quantity, and is therefore unconstrained. In this work, we have used a typical value of 1$^{\circ}$. However, we have also checked with $\theta_c=2^{\circ}$, and found that the upper limit of $\beta$ does not change much.

If the jet aligns with the angular momentum vector of an inclined precessing (due to the LT effect) accretion disc, then a $\beta$ upper limit is also the upper limit of the disc inclination with respect to the black hole equatorial plane. In case the disc inclination angle is equal to the inclination angle ($\beta^*$) of the initial orbit of the disrupted star for J1644+57; our finding implies that $\beta^*$ is less than $\sim 15^{\circ}$. This would have implications for understanding the angular distribution (with respect to the SMBH's spin equatorial plane) of stellar motion near the SMBH's sphere of influence.

We note that, whatever physical reason determines the jet direction, our finding of a small, well-constrained upper limit of jet inclination angle with respect to the black hole spin axis could be useful to probe that physical origin. Such an understanding of the jet physics can be useful not only to study jetted TDEs, but also to probe any jetted accreting system. Finally, we mention that the black hole spin axis could itself be inclined with respect to a plausible black hole spin-precession axis, and in such a case the inferred upper limit of the jet inclination angle could be with respect to the latter axis.

\section*{Acknowledgements}

We thank Srimanta Banerjee for his valuable suggestions and insightful comments on the manuscript, and the Swift Science Data Centre for the data. We also thank an anonymous referee for valuable comments and suggestions for the improvement of this work. C.C. gratefully acknowledges support from the National Key R$\&$D Program of China (Grant No. 2016YFA0400703) and the China Postdoctoral Science Foundation (Grant No. 2018M630023).




\bibliographystyle{mnras}
\bibliography{j1644_lt_tde_2} 

\begin{thebibliography}{}
\makeatletter
\relax
\def\mn@urlcharsother{\let\do\@makeother \do\$\do\&\do\#\do\^\do\_\do\%\do\~}
\def\mn@doi{\begingroup\mn@urlcharsother \@ifnextchar [ {\mn@doi@}
  {\mn@doi@[]}}
\def\mn@doi@[#1]#2{\def\@tempa{#1}\ifx\@tempa\@empty \href
  {http://dx.doi.org/#2} {doi:#2}\else \href {http://dx.doi.org/#2} {#1}\fi
  \endgroup}
\def\mn@eprint#1#2{\mn@eprint@#1:#2::\@nil}
\def\mn@eprint@arXiv#1{\href {http://arxiv.org/abs/#1} {{\tt arXiv:#1}}}
\def\mn@eprint@dblp#1{\href {http://dblp.uni-trier.de/rec/bibtex/#1.xml}
  {dblp:#1}}
\def\mn@eprint@#1:#2:#3:#4\@nil{\def\@tempa {#1}\def\@tempb {#2}\def\@tempc
  {#3}\ifx \@tempc \@empty \let \@tempc \@tempb \let \@tempb \@tempa \fi \ifx
  \@tempb \@empty \def\@tempb {arXiv}\fi \@ifundefined
  {mn@eprint@\@tempb}{\@tempb:\@tempc}{\expandafter \expandafter \csname
  mn@eprint@\@tempb\endcsname \expandafter{\@tempc}}}

\bibitem[\protect\citeauthoryear{{Auchettl}, {Guillochon}  \&
  {Ramirez-Ruiz}}{{Auchettl} et~al.}{2017}]{Auchettl_2017}
{Auchettl} K.,  {Guillochon} J.,   {Ramirez-Ruiz} E.,  2017, \mn@doi [\apj]
  {10.3847/1538-4357/aa633b}, \href
  {http://adsabs.harvard.edu/abs/2017ApJ...838..149A} {838, 149}

\bibitem[\protect\citeauthoryear{{Ayal}, {Livio}  \& {Piran}}{{Ayal}
  et~al.}{2000}]{Ayal_2000}
{Ayal} S.,  {Livio} M.,   {Piran} T.,  2000, \mn@doi [\apj] {10.1086/317835},
  \href {http://adsabs.harvard.edu/abs/2000ApJ...545..772A} {545, 772}

\bibitem[\protect\citeauthoryear{{Bardeen} \& {Petterson}}{{Bardeen} \&
  {Petterson}}{1975}]{Bardeen_Petterson_1975}
{Bardeen} J.~M.,  {Petterson} J.~A.,  1975, \mn@doi [\apjl] {10.1086/181711},
  \href {http://adsabs.harvard.edu/abs/1975ApJ...195L..65B} {195, L65}

\bibitem[\protect\citeauthoryear{{Bardeen}, {Press}  \& {Teukolsky}}{{Bardeen}
  et~al.}{1972}]{Bardeen_1972}
{Bardeen} J.~M.,  {Press} W.~H.,   {Teukolsky} S.~A.,  1972, \mn@doi [\apj]
  {10.1086/151796}, \href {http://adsabs.harvard.edu/abs/1972ApJ...178..347B}
  {178, 347}

\bibitem[\protect\citeauthoryear{{Blandford} \& {Znajek}}{{Blandford} \&
  {Znajek}}{1977}]{Blandford_Znajek_1977}
{Blandford} R.~D.,  {Znajek} R.~L.,  1977, \mn@doi [\mnras]
  {10.1093/mnras/179.3.433}, \href
  {http://adsabs.harvard.edu/abs/1977MNRAS.179..433B} {179, 433}

\bibitem[\protect\citeauthoryear{{Bloom} et~al.,}{{Bloom}
  et~al.}{2011}]{Bloom_2011}
{Bloom} J.~S.,  et~al., 2011, \mn@doi [Science] {10.1126/science.1207150},
  \href {http://adsabs.harvard.edu/abs/2011Sci...333..203B} {333, 203}

\bibitem[\protect\citeauthoryear{{Burrows} et~al.,}{{Burrows}
  et~al.}{2011}]{Burrows_nature_2011}
{Burrows} D.~N.,  et~al., 2011, \mn@doi [\nat] {10.1038/nature10374}, \href
  {http://adsabs.harvard.edu/abs/2011Natur.476..421B} {476, 421}

\bibitem[\protect\citeauthoryear{{Chakraborty} \&
  {Bhattacharyya}}{{Chakraborty} \&
  {Bhattacharyya}}{2017}]{Chakraborty_Bhattacharyya_2017}
{Chakraborty} C.,  {Bhattacharyya} S.,  2017, \mn@doi [\mnras]
  {10.1093/mnras/stx1088}, \href
  {http://adsabs.harvard.edu/abs/2017MNRAS.469.3062C} {469, 3062}

\bibitem[\protect\citeauthoryear{{Chakraborty}, {Kocherlakota}, {Patil},
  {Bhattacharyya}, {Joshi}  \& {Kr{\'o}lak}}{{Chakraborty}
  et~al.}{2017}]{Chakraborty_Kocherlakota_2017}
{Chakraborty} C.,  {Kocherlakota} P.,  {Patil} M.,  {Bhattacharyya} S.,
  {Joshi} P.~S.,   {Kr{\'o}lak} A.,  2017, \mn@doi [\prd]
  {10.1103/PhysRevD.95.084024}, \href
  {http://adsabs.harvard.edu/abs/2017PhRvD..95h4024C} {95, 084024}

\bibitem[\protect\citeauthoryear{{Demircan} \& {Kahraman}}{{Demircan} \&
  {Kahraman}}{1991}]{Demircan-Kahraman_1991}
{Demircan} O.,  {Kahraman} G.,  1991, \mn@doi [\apss] {10.1007/BF00639097},
  \href {http://adsabs.harvard.edu/abs/1991Ap%26SS.181..313D} {181, 313}

\bibitem[\protect\citeauthoryear{{Fragile}}{{Fragile}}{2008}]{Fragile_2008mqw}
{Fragile} P.~C.,  2008, in Microquasars and Beyond. p.~39 (\mn@eprint {arXiv}
  {0810.0526})

\bibitem[\protect\citeauthoryear{{Fragile}, {Blaes}, {Anninos}  \&
  {Salmonson}}{{Fragile} et~al.}{2007}]{Fragile_2007}
{Fragile} P.~C.,  {Blaes} O.~M.,  {Anninos} P.,   {Salmonson} J.~D.,  2007,
  \mn@doi [\apj] {10.1086/521092}, \href
  {http://adsabs.harvard.edu/abs/2007ApJ...668..417F} {668, 417}

\bibitem[\protect\citeauthoryear{{Granot}, {Panaitescu}, {Kumar}  \&
  {Woosley}}{{Granot} et~al.}{2002}]{Granot_2002}
{Granot} J.,  {Panaitescu} A.,  {Kumar} P.,   {Woosley} S.~E.,  2002, \mn@doi
  [\apjl] {10.1086/340991}, \href
  {http://adsabs.harvard.edu/abs/2002ApJ...570L..61G} {570, L61}

\bibitem[\protect\citeauthoryear{{Guillochon} \& {Ramirez-Ruiz}}{{Guillochon}
  \& {Ramirez-Ruiz}}{2013}]{Guillochon_2013}
{Guillochon} J.,  {Ramirez-Ruiz} E.,  2013, \mn@doi [\apj]
  {10.1088/0004-637X/767/1/25}, \href
  {http://adsabs.harvard.edu/abs/2013ApJ...767...25G} {767, 25}

\bibitem[\protect\citeauthoryear{{Kato}}{{Kato}}{1990}]{Kato_1990}
{Kato} S.,  1990, \pasj, \href
  {http://adsabs.harvard.edu/abs/1990PASJ...42...99K} {42, 99}

\bibitem[\protect\citeauthoryear{{Kumar} \& {Granot}}{{Kumar} \&
  {Granot}}{2003}]{Kumar_granot_2003}
{Kumar} P.,  {Granot} J.,  2003, \mn@doi [\apj] {10.1086/375186}, \href
  {http://adsabs.harvard.edu/abs/2003ApJ...591.1075K} {591, 1075}

\bibitem[\protect\citeauthoryear{{Lei}, {Zhang}  \& {Gao}}{{Lei}
  et~al.}{2013}]{Lei_2013}
{Lei} W.-H.,  {Zhang} B.,   {Gao} H.,  2013, \mn@doi [\apj]
  {10.1088/0004-637X/762/2/98}, \href
  {http://adsabs.harvard.edu/abs/2013ApJ...762...98L} {762, 98}

\bibitem[\protect\citeauthoryear{Levan et~al.,}{Levan
  et~al.}{2011}]{Levan_2011}
Levan A.~J.,  et~al., 2011, \mn@doi [Science] {10.1126/science.1207143}, 333,
  199

\bibitem[\protect\citeauthoryear{{Liu} \& {Melia}}{{Liu} \&
  {Melia}}{2002}]{Liu_Melia_2002}
{Liu} S.,  {Melia} F.,  2002, \mn@doi [\apjl] {10.1086/341991}, \href
  {http://adsabs.harvard.edu/abs/2002ApJ...573L..23L} {573, L23}

\bibitem[\protect\citeauthoryear{{Lodato} \& {Rossi}}{{Lodato} \&
  {Rossi}}{2011}]{Lodato_2011}
{Lodato} G.,  {Rossi} E.~M.,  2011, \mn@doi [\mnras]
  {10.1111/j.1365-2966.2010.17448.x}, \href
  {http://adsabs.harvard.edu/abs/2011MNRAS.410..359L} {410, 359}

\bibitem[\protect\citeauthoryear{{Lodato}, {King}  \& {Pringle}}{{Lodato}
  et~al.}{2009}]{Lodato_2009}
{Lodato} G.,  {King} A.~R.,   {Pringle} J.~E.,  2009, \mn@doi [\mnras]
  {10.1111/j.1365-2966.2008.14049.x}, \href
  {http://adsabs.harvard.edu/abs/2009MNRAS.392..332L} {392, 332}

\bibitem[\protect\citeauthoryear{{Lu}, {Krolik}, {Crumley}  \& {Kumar}}{{Lu}
  et~al.}{2017}]{Lu_2017}
{Lu} W.,  {Krolik} J.,  {Crumley} P.,   {Kumar} P.,  2017, \mn@doi [\mnras]
  {10.1093/mnras/stx1668}, \href
  {http://adsabs.harvard.edu/abs/2017MNRAS.471.1141L} {471, 1141}

\bibitem[\protect\citeauthoryear{{Mangalam} \& {Mageshwaran}}{{Mangalam} \&
  {Mageshwaran}}{2018}]{Mangalam_TDE_2018}
{Mangalam} A.,  {Mageshwaran} T.,  2018, Bulletin de la Societe Royale des
  Sciences de Liege, \href {http://adsabs.harvard.edu/abs/2018BSRSL..87..307M}
  {87}

\bibitem[\protect\citeauthoryear{{Mangano}, {Burrows}, {Sbarufatti}  \&
  {Cannizzo}}{{Mangano} et~al.}{2016}]{Mangano_2016}
{Mangano} V.,  {Burrows} D.~N.,  {Sbarufatti} B.,   {Cannizzo} J.~K.,  2016,
  \mn@doi [\apj] {10.3847/0004-637X/817/2/103}, \href
  {http://adsabs.harvard.edu/abs/2016ApJ...817..103M} {817, 103}

\bibitem[\protect\citeauthoryear{{McKinney}}{{McKinney}}{2006}]{McKinney_2006}
{McKinney} J.~C.,  2006, \mn@doi [\mnras] {10.1111/j.1365-2966.2006.10256.x},
  \href {http://adsabs.harvard.edu/abs/2006MNRAS.368.1561M} {368, 1561}

\bibitem[\protect\citeauthoryear{{McKinney}, {Tchekhovskoy}  \&
  {Blandford}}{{McKinney} et~al.}{2013}]{McKinney_2013}
{McKinney} J.~C.,  {Tchekhovskoy} A.,   {Blandford} R.~D.,  2013, \mn@doi
  [Science] {10.1126/science.1230811}, \href
  {http://adsabs.harvard.edu/abs/2013Sci...339...49M} {339, 49}

\bibitem[\protect\citeauthoryear{Meier, Koide  \& Uchida}{Meier
  et~al.}{2001}]{Meier_2001}
Meier D.~L.,  Koide S.,   Uchida Y.,  2001, \mn@doi [Science]
  {10.1126/science.291.5501.84}, 291, 84

\bibitem[\protect\citeauthoryear{{Natarajan} \& {Pringle}}{{Natarajan} \&
  {Pringle}}{1998}]{Natarajan_1998}
{Natarajan} P.,  {Pringle} J.~E.,  1998, \mn@doi [\apjl] {10.1086/311658},
  \href {http://adsabs.harvard.edu/abs/1998ApJ...506L..97N} {506, L97}

\bibitem[\protect\citeauthoryear{{Nixon} \& {King}}{{Nixon} \&
  {King}}{2013}]{Nixon_2013}
{Nixon} C.,  {King} A.,  2013, \mn@doi [\apjl] {10.1088/2041-8205/765/1/L7},
  \href {http://adsabs.harvard.edu/abs/2013ApJ...765L...7N} {765, L7}

\bibitem[\protect\citeauthoryear{{Phinney}}{{Phinney}}{1989}]{Phinney_1989}
{Phinney} E.~S.,  1989, in {Morris} M.,  ed.,  IAU Symposium Vol. 136, The
  Center of the Galaxy. p.~543

\bibitem[\protect\citeauthoryear{{Rees}}{{Rees}}{1988}]{Rees_1988Nature}
{Rees} M.~J.,  1988, \mn@doi [\nat] {10.1038/333523a0}, \href
  {http://adsabs.harvard.edu/abs/1988Natur.333..523R} {333, 523}

\bibitem[\protect\citeauthoryear{{Reis} \& {Miller}}{{Reis} \&
  {Miller}}{2013}]{Reis_2013}
{Reis} R.~C.,  {Miller} J.~M.,  2013, \mn@doi [\apjl]
  {10.1088/2041-8205/769/1/L7}, \href
  {http://adsabs.harvard.edu/abs/2013ApJ...769L...7R} {769, L7}

\bibitem[\protect\citeauthoryear{{Salafia}, {Ghisellini}, {Pescalli},
  {Ghirlanda}  \& {Nappo}}{{Salafia} et~al.}{2015}]{salafia_2015}
{Salafia} O.~S.,  {Ghisellini} G.,  {Pescalli} A.,  {Ghirlanda} G.,   {Nappo}
  F.,  2015, \mn@doi [\mnras] {10.1093/mnras/stv766}, \href
  {http://adsabs.harvard.edu/abs/2015MNRAS.450.3549S} {450, 3549}

\bibitem[\protect\citeauthoryear{{Saxton}, {Soria}, {Wu}  \& {Kuin}}{{Saxton}
  et~al.}{2012}]{Saxton_2012}
{Saxton} C.~J.,  {Soria} R.,  {Wu} K.,   {Kuin} N.~P.~M.,  2012, \mn@doi
  [\mnras] {10.1111/j.1365-2966.2012.20739.x}, \href
  {http://adsabs.harvard.edu/abs/2012MNRAS.422.1625S} {422, 1625}

\bibitem[\protect\citeauthoryear{{Shen} \& {Matzner}}{{Shen} \&
  {Matzner}}{2012}]{Shen_Matzner_2012}
{Shen} R.-F.,  {Matzner} C.~D.,  2012, in European Physical Journal Web of
  Conferences. p. 07006, \mn@doi{10.1051/epjconf/20123907006}

\bibitem[\protect\citeauthoryear{{Shen} \& {Matzner}}{{Shen} \&
  {Matzner}}{2014}]{Shen_Matzner_2014}
{Shen} R.-F.,  {Matzner} C.~D.,  2014, \mn@doi [\apj]
  {10.1088/0004-637X/784/2/87}, \href
  {http://adsabs.harvard.edu/abs/2014ApJ...784...87S} {784, 87}

\bibitem[\protect\citeauthoryear{{Stone} \& {Loeb}}{{Stone} \&
  {Loeb}}{2012}]{SL_2012}
{Stone} N.,  {Loeb} A.,  2012, \mn@doi [Physical Review Letters]
  {10.1103/PhysRevLett.108.061302}, \href
  {http://adsabs.harvard.edu/abs/2012PhRvL.108f1302S} {108, 061302}

\bibitem[\protect\citeauthoryear{{Strubbe} \& {Quataert}}{{Strubbe} \&
  {Quataert}}{2009}]{Strubbe_Quataert_2009}
{Strubbe} L.~E.,  {Quataert} E.,  2009, \mn@doi [\mnras]
  {10.1111/j.1365-2966.2009.15599.x}, \href
  {http://adsabs.harvard.edu/abs/2009MNRAS.400.2070S} {400, 2070}

\bibitem[\protect\citeauthoryear{{Tchekhovskoy}, {Metzger}, {Giannios}  \&
  {Kelley}}{{Tchekhovskoy} et~al.}{2014}]{Tchekhovskoy_2014}
{Tchekhovskoy} A.,  {Metzger} B.~D.,  {Giannios} D.,   {Kelley} L.~Z.,  2014,
  \mn@doi [\mnras] {10.1093/mnras/stt2085}, \href
  {http://adsabs.harvard.edu/abs/2014MNRAS.437.2744T} {437, 2744}

\bibitem[\protect\citeauthoryear{{Ulmer}}{{Ulmer}}{1999}]{Ulmer_1999}
{Ulmer} A.,  1999, \mn@doi [\apj] {10.1086/306909}, \href
  {http://adsabs.harvard.edu/abs/1999ApJ...514..180U} {514, 180}

\bibitem[\protect\citeauthoryear{{Wang}, {Lei}, {Wang}, {Zou}, {Zhang}, {Gao}
  \& {Huang}}{{Wang} et~al.}{2014}]{Wang_2014}
{Wang} J.-Z.,  {Lei} W.-H.,  {Wang} D.-X.,  {Zou} Y.-C.,  {Zhang} B.,  {Gao}
  H.,   {Huang} C.-Y.,  2014, \mn@doi [\apj] {10.1088/0004-637X/788/1/32},
  \href {http://adsabs.harvard.edu/abs/2014ApJ...788...32W} {788, 32}

\bibitem[\protect\citeauthoryear{{Wang}, {Lei}, {Wang}, {Xie}  \&
  {Zhang}}{{Wang} et~al.}{2015}]{Wang_2015}
{Wang} J.,  {Lei} W.,  {Wang} D.,  {Xie} W.,   {Zhang} B.,  2015, preprint,
  \href {http://adsabs.harvard.edu/abs/2015arXiv151102488W} {} (\mn@eprint
  {arXiv} {1511.02488})

\bibitem[\protect\citeauthoryear{{Zauderer} et~al.,}{{Zauderer}
  et~al.}{2011}]{Zauderer_nature_2011}
{Zauderer} B.~A.,  et~al., 2011, \mn@doi [\nat] {10.1038/nature10366}, \href
  {http://adsabs.harvard.edu/abs/2011Natur.476..425Z} {476, 425}

\bibitem[\protect\citeauthoryear{{Zauderer}, {Berger}, {Margutti}, {Pooley},
  {Sari}, {Soderberg}, {Brunthaler}  \& {Bietenholz}}{{Zauderer}
  et~al.}{2013}]{Zauderer_2013}
{Zauderer} B.~A.,  {Berger} E.,  {Margutti} R.,  {Pooley} G.~G.,  {Sari} R.,
  {Soderberg} A.~M.,  {Brunthaler} A.,   {Bietenholz} M.~F.,  2013, \mn@doi
  [\apj] {10.1088/0004-637X/767/2/152}, \href
  {http://adsabs.harvard.edu/abs/2013ApJ...767..152Z} {767, 152}

\makeatother
\end{thebibliography}

\bsp	
\label{lastpage}
\end{document}